# Generation and characterization of coherent terahertz radiation from 100-TW laser-wakefield acceleration

Taegyu Pak[1,2,3], Dae Hee Wi[1,2], Sang Beom Kim[1,2], Jaewon Lim[1,2], Jae Hee Sung[1,4], Seong Ku Lee[1,4], and Ki-Yong Kim[1,2,5†]

[1]Center for Relativistic Laser Science, Institute for Basic Science, Gwangju 61005, Korea
[2]Department of Physics and Photon Science, Gwangju Institute of Science and Technology, Gwangju 61005, Korea
[3]Deutsches Elektronen-Synchrotron DESY, Notkestrasse 85, Hamburg 22607, Germany
[4]Advanced Photonics Research Institute, Gwangju Institute of Science and Technology, Gwangju 61005, Korea
[5]Present address: Institute for Research in Electronics and Applied Physics; Department of Physics, University of Maryland, College Park, Maryland 20742, USA
[†]kykim@umd.edu

**Abstract**
We experimentally characterized terahertz (THz) radiation emitted from laser-wakefield acceleration (LWFA) driven at 100-TW laser power. Simultaneous measurements of the laser energy, electron-bunch charge, and THz energy reveal a quadratic dependence of the THz energy on both charge and laser energy. This behavior indicates coherent collective emission in the generation process and provides a useful scaling law for THz output. Microbolometer-based beam profiling shows a relatively large THz beam divergence (~0.2 rad). Single-shot THz interferometry further shows that the emitted THz pulse is sub-picosecond in duration and broadband. Combining the beam-profile and interferometric measurements, the THz spectrum is expected to span approximately 1–20 THz. Together, these results support coherent acceleration radiation as the dominant mechanism for THz generation in 100-TW LWFA

## 1. Introduction

The terahertz (THz) band of the electromagnetic spectrum (1 THz = $10^{12}$ Hz)—bridging the microwave and infrared domains—has traditionally been regarded as a particularly challenging frontier in photonics. However, rapid progress in THz source and detector technologies is transforming the field, making the THz range increasingly accessible for both fundamental studies and practical applications in spectroscopy, imaging, sensing, and communications [1-3]. In particular, the emergence of strong-field THz sources is opening access to new regimes of light-matter interaction, enabling THz-driven nonlinear spectroscopy [4-6], ultrafast phase transitions [7], high-harmonic generation [8, 9], charged-particle acceleration [10], and THz-induced tunneling ionization [11].

All of these emerging applications rely on strong THz sources, and a variety of high-energy THz sources have been developed. These include difference-frequency generation [12-14] and optical rectification [11, 15-18] of ultrafast laser pulses in nonlinear crystals, single- and two-color laser mixing in gases [19-24], coherent transition radiation (CTR) from ultrashort electron bunches [25], and laser-plasma interactions in liquids [26, 27] and solids [28-30]. Notably, tens of millijoules of THz energy were obtained from metal foils irradiated with ~60 J picosecond laser pulses [29], and joule-level THz radiation has been produced from petawatt (PW) laser-solid interactions [30].

Laser-wakefield acceleration (LWFA) [31] offers another promising route to high-energy THz generation [32-37]. Relativistic electron bunches, produced by LWFA, emit THz radiation as they exit the plasma-vacuum boundary via CTR [32-35]. In general, coherent THz emission occurs when the source (electron bunch) dimensions are comparable to or smaller than the emitted THz wavelength; under this condition, the fields radiated by individual electrons add coherently, and the radiated energy scales with the square of the total charge.

Early experiments with 10-TW-class lasers produced sub-100 nJ THz pulses, consistent with the CTR model [32-35]. However, a recent experiment at 100-TW laser power generated ~4 mJ of THz radiation; such a high THz output cannot be explained by the conventional CTR model alone, and a coherent acceleration radiation (CAR) model was proposed to account for this regime [36]. In the CAR model, THz waves are produced by relatively low-energy (~MeV) but high-charge (~nC) electrons undergoing strong acceleration and scattering (wave breaking) near the back of plasma bubbles. This occurs continuously over the entire plasma length, producing coherent, conical, radially polarized broadband THz emission [36, 37].

Beyond the CTR and CAR models, additional mechanisms have been proposed to explain or predict THz radiation in conventional LWFA setups. In linear mode conversion (LMC) [38, 39], longitudinal wakefields couple to transverse electromagnetic modes in a density gradient, producing narrowband THz radiation near the plasma frequency, often emitted backward. Cherenkov-type radiation (CR) [40, 41] occurs when superluminal wake structures in an up-ramp radiate into propagating electromagnetic waves, yielding conical, radially polarized emission with narrowband components and harmonics of the plasma frequency. Photon deceleration (PD) [42, 43] arises from wakefield-induced redshifting of photons in density-tailored plasmas, converting optical energy into lower-frequency radiation that can be broadband and extend toward high-frequency THz/mid-IR.

All of these mechanisms arise in LWFA-based setups, but they differ in their underlying physics and therefore predict distinct THz properties. Accordingly, comprehensive characterization of the THz radiation is essential for identifying the dominant THz generation mechanism(s). Our previous work investigated high-energy THz generation under various conditions—including laser focal position, gas species, and plasma density—and found that the THz output energy is strongly correlated with the electron-beam charge, rather than necessarily with the electron-beam energy. We also confirmed that the emitted THz radiation is radially polarized. However, THz beam profiles were not measured, and no direct test was performed to verify whether the radiation is coherent. In addition, the spectrum was only coarsely characterized, providing no information on the THz pulse duration.

In this work, we present a follow-up experimental investigation characterizing THz radiation emitted at 100-TW-level LWFA. We implement three complementary diagnostics: (i) simultaneous measurements of laser energy, electron-bunch charge, and THz energy to quantify their correlations; (ii) THz beam profiling and divergence characterization; and (iii) single-shot THz interferometry to retrieve temporal and spectral information. Together, these measurements demonstrate that the generated THz radiation is coherent, highly divergent, sub-picosecond in duration, and broadband.

## 2. Experimental setup

A 100-TW Ti:sapphire laser system [44] (800 nm central wavelength, 27 fs pulse duration, <2.7 J, linear polarization) was used to drive LWFA and simultaneously generate relativistic electron beams with intense THz radiation. The laser was focused by a 1.5-m focal-length concave mirror to a 22-µm full width at half maximum (FWHM) spot, yielding a peak intensity of approximately $5.2\times10^{18}$ W/cm² in vacuum [36].

Figure 1(a) shows a schematic of the LWFA interaction vacuum chamber. The 100-TW laser pulse was focused into a nitrogen gas jet, where relativistic electrons and intense THz radiation were simultaneously generated. A solenoid-valve gas jet (Parker, Series 9) equipped with a 4-mm-diameter cylindrical nozzle supplied the target gas. The jet was operated in pulsed mode with a 10-ms opening time. The plasma density profile was characterized using a wavefront sensor (Phasics, SID4-HR) by measuring the phase modulation induced on a separate probe beam. For gas backing pressures of 7–28 bar, the on-axis gas density ranged from 0.4–$1.5\times10^{18}$ cm$^{-3}$ at a height of 2 mm above the nozzle, exhibiting a ~2-mm-wide plateau surrounded by 1-mm up- and down-ramps [36]. Relativistic electrons produced by LWFA were diagnosed using two scintillating screens (Kodak, Lanex). Lanex 1 was placed 320 mm downstream of the gas jet at a 45° angle to measure the electron beam profile, pointing, divergence, and charge. The electron energy distribution was obtained using a 1-T dipole-magnet spectrometer (205 mm × 70 mm × 8 mm gap), followed by Lanex 2 located 536 mm from the magnet exit, covering energies of 20–350 MeV. All Lanex screens were imaged with 16-bit, 2560×2160 sCMOS cameras (PCO Imaging, PCO.edge 5.5 USB sCMOS camera).

Figure 1(b) illustrates the THz collection and transport optics installed downstream of the LWFA interaction chamber. THz radiation emitted from the plasma was first collected by a holed, gold-coated, off-axis parabolic (OAP) mirror with a 50-mm diameter, 102-mm focal length, and a 12-mm-diameter central hole. This configuration provided a half-collection angle of approximately 3.4°–14°. The high-power laser pulse and relativistic electron beam propagated through the central hole, while the THz radiation was reflected by the OAP mirror. After reflection, the THz beam exited the vacuum chamber through a 180-µm-thick Mylar window, preceded by a high-density polyethylene (HDPE) filter that attenuates infrared leakage but transmits THz radiation efficiently. Outside the chamber, the THz beam was refocused by a second OAP mirror (60-mm clear aperture, 230-mm focal length) and subsequently guided to three diagnostics—a pyroelectric detector (PED; Gentec-EO, THZ5I-BL-BNC) for energy measurements, a microbolometer camera (FLIR, A65) for beam profiling, and a single-shot interferometer for temporal and spectral characterization [45]. The THz beam passed through different combinations of optical elements, including a THz beamsplitter (TYDEX, BS-HRFZ-SI-D76.2-T6), a THz attenuator (TYDEX, ATS-5-50.8), and a low-pass filter (TYDEX, LPF 23.1-47) depending on the diagnostic configuration, as shown in Fig. 1(b).

Representative diagnostic outputs are summarized in Fig. 1(c–g). Figure 1(c) shows an LWFA electron-beam profile obtained from Lanex 1, where low-energy electrons exhibit a relatively large divergence. Figure 1(d) presents the corresponding electron-energy spectrum from the magnetic spectrometer, with most electrons lying below 20 MeV. We note that the low-energy (~MeV) but high-charge (~nC) electrons that dominate THz generation typically have a large divergence and are therefore largely clipped at the spectrometer entrance or not detected by Lanex 2 because of its limited detection range. Figure 1(e) shows a typical THz signal measured using the PED. Figure 1(f) shows a THz beam profile recorded by the A65 camera; the central dark region originates from the 12-mm hole in the holed OAP mirror used for THz collection. Figure 1(g) displays a single-shot THz interferogram captured by the interferometer's microbolometer camera.

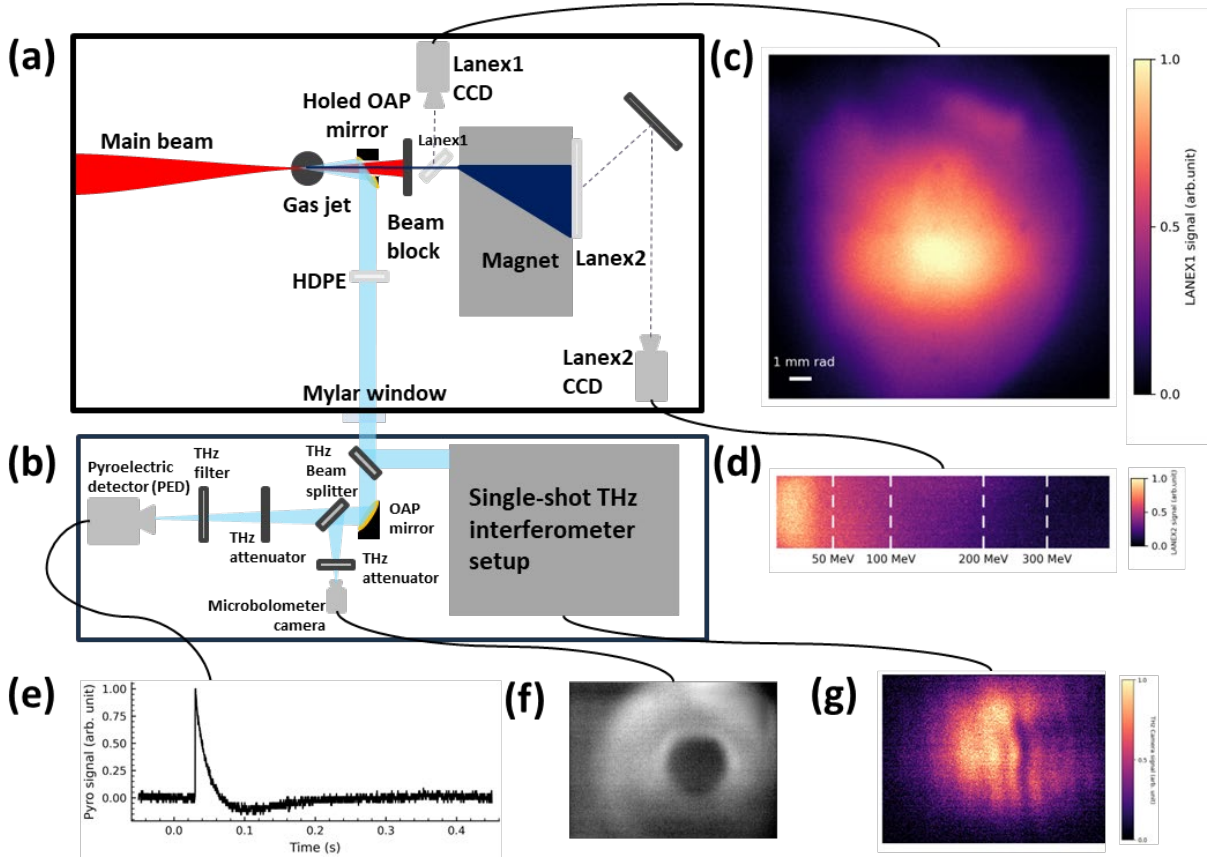

**Fig. 1.** (a) Schematic of the LWFA interaction inside the vacuum chamber: the 150-TW laser pulse is focused into a gas jet to drive LWFA, generating relativistic electron beams and THz radiation. (b) THz diagnostic setup, including a pyroelectric detector (PED), a microbolometer camera, and a single-shot Mach-Zehnder interferometer. (c) Electron-beam profile on Lanex 1 (false color) imaged by the CCD camera. (d) Electron energy spectrum measured from Lanex 2 in the magnetic spectrometer. (e) Representative PED signal recorded on an oscilloscope. (f) THz beam profile recorded by the microbolometer camera; the central dark region arises from the hole in the holed OAP mirror. (g) Single-shot THz interferogram showing interference fringes (false color).

## 3. Results
### 3.1 Energy scaling of the emitted THz radiation

To study the energy scaling of the THz radiation, we simultaneously measured the THz pulse energy using the PED, the laser energy delivered to the LWFA target, and the electron-beam charge measured by Lanex 1. Figures 2(a) and 2(b) summarize the resulting correlations among these quantities. Figure 2(a) shows the output THz energy as a function of the input laser energy. The THz energy exhibits large shot-to-shot fluctuations even at nearly constant laser energy, and the magnitude of these fluctuations increases with increasing laser energy. Despite these variations, the overall trend indicates a nonlinear increase in THz output with laser energy.

To better understand this nonlinear trend, we replotted the THz energy as a function of the electron-beam charge $q$, as shown in Fig. 2(b). The data generally follow a $q^2$ dependence, indicating that the THz radiation is coherently emitted from individual electrons, consistent with the coherent radiation models including CTR and CAR. In LWFA, the laser spot radius $r$ typically scales with laser energy as $r^2 \propto E_L$ to keep the laser peak intensity constant. This leads to a corresponding increase in beam charge, $q \propto n_e c \tau r^2 \propto E_L$, where $n_e$ is the electron density (constant for constant intensity) and $\tau$ is the laser pulse duration (also constant). Therefore, the THz energy is expected to scale as $E_{THz} \propto q^2 \propto E_L^2$. This quadratic scaling agrees well with the trend shown in Fig. 2(a).

We note that the THz energy scaling can be further enhanced by accounting for the plasma length. In our experiment, the plasma length was fixed by the gas-jet length (<4 mm). In principle, the plasma (or gas-jet) length $l$ can be increased with laser energy for efficient LWFA operation, since the Rayleigh length $z_R$ scales as $z_R \propto r^2$ and $l$ is typically chosen to be proportional to $z_R$. This yields $l \propto z_R \propto r^2 \propto E_L$. In addition, the CAR model predicts a linear scaling of the THz energy with plasma length, $E_{THz} \propto l \propto E_L$ [37]. Combining the charge and plasma-length contributions leads to a highly nonlinear overall scaling, $E_{THz} \propto q^2 l \propto E_L^3$, assuming negligible energy depletion. This strong dependence suggests that LWFA at extremely high laser powers—particularly in the petawatt (PW) regime—has the potential to produce coherent THz pulses with energies at the multi-joule level. For example, ~4 mJ of THz energy [36] was produced at 100 TW (2.7 J, 27 fs); the cubic scaling predicts THz energies as high as ~4 J at 1 PW laser power (27 J, 27 fs).

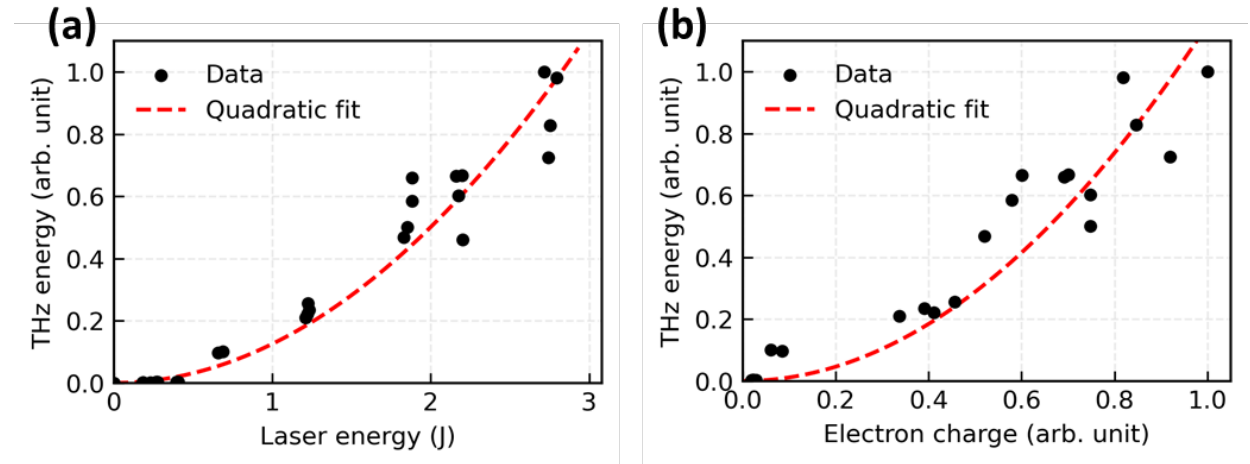

**Fig. 2.** Output THz energy (scatter points) as a function of (a) laser energy and (b) electron-beam charge. In both cases, the data follow quadratic trend lines (red dashed).

### 3.2 THz beam-profile measurements

To investigate the spatial characteristics of the emitted THz beam, we measured the THz beam profile using an uncooled microbolometer camera (FLIR, A65). The overall imaging configuration is schematically shown in Fig. 3(a). The generated THz beam was first collected by the holed OAP mirror and then focused by the second OAP mirror onto the A65 camera. To accurately locate the image plane (camera position), a reference source—a heated soldering-iron tip—was placed at the gas-jet position; the camera position was adjusted to image the front surface of the OAP mirror as the object plane. A sample image, shown in Fig. 3(b), clearly resolves the central aperture and the boundary of the holed OAP mirror. The corresponding THz beam profiles (three separate shots) obtained under LWFA operation at the same camera position are shown in Fig. 3(c).

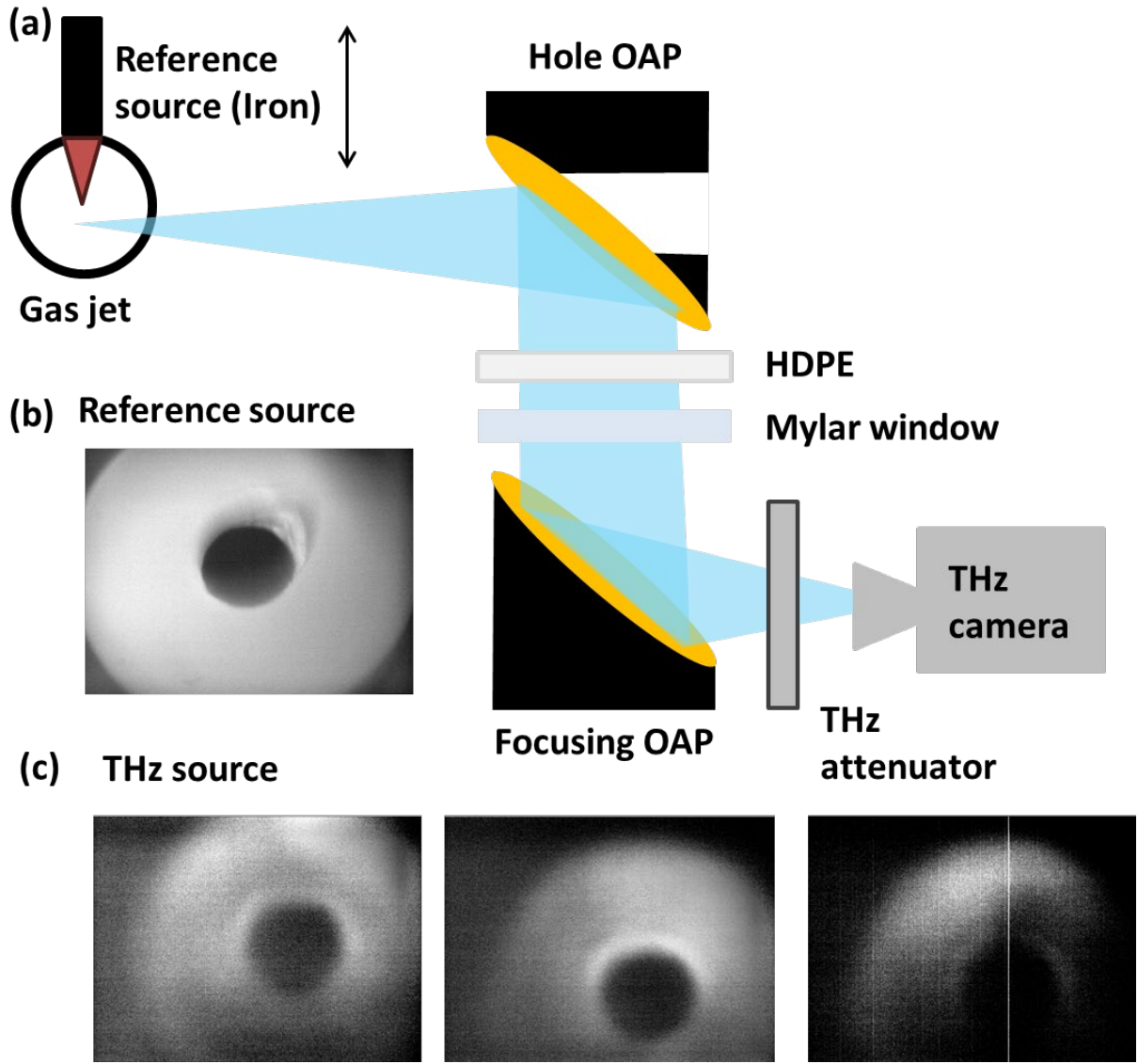

**Fig. 3.** (a) Schematic of the THz beam-imaging setup using a holed OAP mirror and a secondary focusing OAP mirror. (b) Thermal images recorded with a reference source (soldering iron) to identify the imaging plane (c) Corresponding THz beam images measured during LWFA operation at the camera same positions.

The measured THz beam profile exhibits a continuous distribution with a size comparable to that of the OAP mirror, corresponding to a half-divergence angle of ~0.2 rad. The central dark region in each image corresponds to the 12-mm aperture of the holed OAP, which obscures the THz beam profile. As a result, it is difficult to determine whether the beam profile is conical in the far field. Due to shot-to-shot variations, some THz profiles are not perfectly symmetric. This asymmetry may explain the asymmetric polarization pattern observed in our previous measurements [36].

The large THz divergence angle (~0.2 rad) also suggests that the radiation originates from relatively low-energy electrons (~MeV), since radiation emitted by an electron under constant acceleration has a characteristic divergence of ~$1/(2\gamma)$, where $\gamma$ is the relativistic Lorentz factor.

This estimate assumes a point source and does not account for continuous emission along the interaction (plasma) length. Because the effective radiation angle generally decreases as the longitudinal extent of the moving source increases, the measured divergence provides an upper bound on the electron energy responsible for the THz emission.

We note that the A65 camera is primarily designed for long-wavelength infrared (LWIR) detection at 8–15 µm (20–37 THz), but it provides a broader effective detection range of 5–50 THz [46]. This range, however, is constrained by the filters (HDPE, Mylar, beamsplitter, and attenuator) in the beam path that provide a transmission window of 0.1–20 THz (with strong attenuation at 10-18 THz). Combined with the A65 response, the effective detection bandwidth is 5–20 THz. Accordingly, the THz profiles in Fig. 3(c) predominantly represent emission within this frequency range.

### 3.3 Single-shot THz interferometry

To characterize the spectrum of the generated THz radiation, we employed a single-shot THz interferometer. Previously, multiple THz bandpass filters were inserted sequentially in front of the PED to obtain a discrete spectral distribution [36]. However, this multi-shot method is time-consuming and highly vulnerable to shot-to-shot fluctuations. Because LWFA and the accompanying THz generation are highly nonlinear and exhibit substantial fluctuations, single-shot spectral characterization is highly desirable.

Our single-shot diagnostic is based on a Mach-Zehnder THz interferometer [45], as shown in Fig. 4(a). After entering the interferometer, the THz pulse is split into two paths: one with a fixed optical delay and the other with a motorized delay stage. The two beams are then weakly focused by separate OAP mirrors (both with focal lengths of 191 mm) and recombined with a small tilt angle. This tilt maps the temporal autocorrelation onto a spatial coordinate, producing interferometric fringes on the detector upon recombination. A similar configuration was recently used to spectrally characterize multicycle 15-THz radiation generated by optical rectification in lithium niobate [11].

Figure 4(b) shows a representative interferogram recorded by a room-temperature microbolometer camera (Swiss Terahertz, RIGI M2) sensitive in the 0.1–10 THz range. In this setup, the temporal delay $\tau$ is related to the transverse coordinate on the camera sensor ($x$) as $\tau = x\sin\theta/c$, where $\theta$ is the tilt angle and $c$ is the speed of light [45]. Here, the tilt angle was measured to be $\theta = 10.6°$ using an independent thermal source and performing $\tau$–$x$ mapping. This allowed us to convert the spatial interferogram in Fig. 4(b) into a one-dimensional (1D) autocorrelation signal via multi-row integration, as shown in Fig. 4(c). The autocorrelation signal was then Fourier-transformed to retrieve the corresponding spectrum (Fig. 4(d)).

The reconstructed spectrum shows emission at 1–10 THz. The strong signal below 1 THz (red dashed line) is an artifact arising from the THz beam envelope itself (red dashed line in Fig. 4(c)). The spectrum beyond ~6 THz is largely suppressed by the 14.3-THz low-pass filter (TYDEX, LPF 14.3) and additional filters placed before the single-shot diagnostic. Taken together with the emission observed using the A65 camera, the generated THz radiation is expected to extend from 1 to ~20 THz, mostly below the estimated plasma frequency of ~20 THz. Furthermore, the measured temporal autocorrelation indicates a sub-picosecond transform-limited pulse duration for the emitted THz waves.

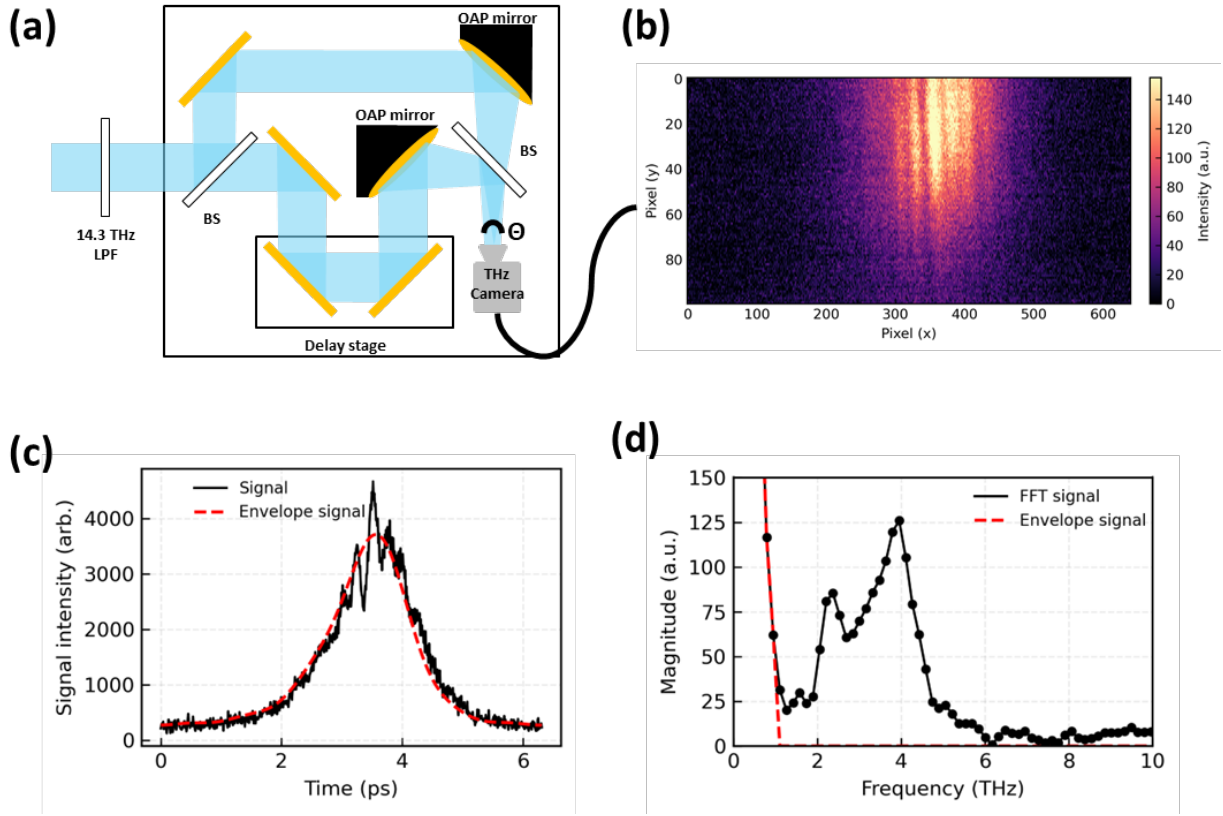

**Fig. 4.** (a) Schematic of our single-shot Mach-Zehnder interferometer used for THz temporal and spectral characterization. (b) Sample interferogram recorded by the THz camera. (c) Temporal autocorrelation signal obtained by integrating the interferogram over pixel rows after converting the spatial coordinate to temporal delay. (d) Retrieved THz spectrum via Fourier transformation, coplotted with the low-frequency artifact (envelope signal) below 1 THz (red line).

## 4. Discussion & Conclusions

Our spatial, temporal, and spectral characterization reveals that the THz radiation produced in 100-TW LWFA is coherent, highly divergent, sub-picosecond in duration, and broadband, spanning approximately 1–20 THz. The THz radiation is also radially polarized, as previously reported [36].

The observed quadratic dependence of THz energy on bunch charge indicates coherent emission from the electron beam, supporting electron-beam-based models such as CTR and CAR. The measured spectrum (1–20 THz), which lies predominantly below the plasma frequency (~20 THz), also disfavors LMC, CR, and PD as dominant mechanisms under our experimental conditions. Moreover, LMC is unlikely in our case because the detected radiation is forward-directed.

Because CTR and CAR share many characteristics—coherence, radial polarization, sub-picosecond duration, and broadband spectra—distinguishing between them is not straightforward. Nonetheless, the large THz divergence suggests that the radiation originates primarily from low-energy (~MeV) electrons, rather than from the high-energy (>20 MeV) electron bunch accelerated within the wake. In addition, the charge of the high-energy electron bunch is much smaller than that of the low-energy electron population and is therefore insufficient to account for the observed THz energy. It remains possible that low-energy electrons continue propagating forward and generate CTR at the plasma-vacuum boundary; however, this scenario would imply a substantially enlarged beam size (hundreds of microns) at the boundary, which would not support coherent emission up to 20 THz (corresponding to a maximum effective source size of <15 μm). This coherence requirement is more naturally satisfied by CAR, in which radiation is produced near wave-breaking points behind plasma bubbles. A more rigorous test to identify the dominant mechanism(s) would be to vary the plasma length and directly measure the resulting THz-energy scaling, since CAR arises from a continuous, line-like source along the interaction region, whereas CTR occurs at a localized region (plasma–vacuum boundary). In addition, such a length-dependence study would provide a direct test of the cubic scaling law for THz output, potentially paving the way toward coherent multi-joule THz generation in PW-level LWFA.

Finally, the diagnostic framework introduced here—energy scaling, beam profiling, and single-shot interferometry—can be readily extended to investigate THz emission under different plasma conditions, injection schemes, or laser parameters. This approach provides a pathway to distinguish among competing THz-generation mechanisms in LWFA and ultimately to guide the development of higher-energy and higher-repetition-rate THz sources.